\documentclass[aps,prapplied,superscriptaddress,twocolumn]{revtex4-2}
\UseRawInputEncoding
\usepackage[colorlinks=true, allcolors=blue]{hyperref}
\usepackage{xurl}

\usepackage{amsmath}

\usepackage{graphicx}
\usepackage{multirow}
\usepackage{lineno}
\usepackage{xcolor}
\usepackage{booktabs}
\usepackage{hyperref}

\begin{document}

%\preprint{}

\title{Entanglement distribution over metropolitan fiber using on-chip broadband polarization entangled photon source} 



\author{Zi-Heng Jiang}
\author{Yikai Chen}
\author{Wenhan Yan}
\author{Chi Lu}
\author{Wenjun Wen}
\author{Yu-Yang An}
\author{Leizhen Chen}
\author{Yuchen Liu}
\author{Hua-Ying Liu}
\author{Zhenda Xie}
\author{Yan-Qing Lu}
\author{Shining Zhu}
\affiliation{National Laboratory of Solid-State Microstructures, School of Physics, College of Engineering and Applied Sciences, School of Electronic Science and Engineering, Collaborative Innovation Center of Advanced Microstructures, Jiangsu Key Laboratory of Quantum Information Science and Technology, Nanjing University, Nanjing 210093,China}
\author{Xiao-Song Ma}
\email{Xiaosong.Ma@nju.edu.cn}
\affiliation{National Laboratory of Solid-State Microstructures, School of Physics, College of Engineering and Applied Sciences, School of Electronic Science and Engineering, Collaborative Innovation Center of Advanced Microstructures, Jiangsu Key Laboratory of Quantum Information Science and Technology, Nanjing University, Nanjing 210093,China}
\affiliation{Synergetic Innovation Center of Quantum Information and Quantum Physics, University of Science and Technology of China, Hefei, Anhui 230026, China}
\affiliation{Hefei National Laboratory, Hefei 230088, China}

%\thanks{}


\date{\today}

\begin{abstract}
Entangled photon pairs are of crucial importance in quantum networks. For the future demands of large-scale and secure quantum communication, integrated photon sources are highly effective solutions. Here, we report entanglement distribution over a 30 km metropolitan area using on-chip broadband silicon nanowire biphoton polarization entangled source based on a silicon-on-insulator (SOI) platform. This source generates a continuous spectrum spanning the entire C-band (4.5 THz), achieving a locally detected coincidence counts of about 154 kHz within 100 GHz bandwidth, making it suitable for long-distance entanglement distribution among multiple users. By combining this source with quantum entanglement, enhanced by high-precision clock synchronization that achieves a standard deviation of 56.8 ps over 600 s, we observe a violation of the CHSH inequality by 27.8 standard deviations. Our results showcase the potential of silicon photonic technology as a scalable and practical platform for quantum technologies.
\end{abstract}

%\keywords{}

\maketitle

\section{Introduction}
%\subsection{}
%\subsubsection{}

Efficient entanglement sources play a crucial role in the construction of quantum networks, enabling applications such as quantum communication\cite{gisin2007quantum,ursin2007entanglement,yin2012quantum,ma2012quantum,ren2017ground,chen2021integrated,lu2022micius} and quantum computing\cite{nielsen2010quantum,ladd2010quantum}. Fiber-based entanglement distribution systems offer compatibility with existing telecommunications infrastructure. In recent years,significant progress has been made in the development of large-scale fiber quantum networks\cite{lucamarini2018overcoming,wengerowsky2019entanglement,dynes2019cambridge,wengerowsky2020passively,chen2021twin,neumann2022continuous,wang2022twin,liu2023experimental}.These networks primarily utilize traditional bulk optics for realizing the transmitters or receivers.

Silicon-based integrated photonic chips serve as reliable platforms for implementing optical quantum technologies\cite{takesue2008generation,leuthold2010nonlinear,silverstone2014chip,zhang2019generation,feng2019generation,paesani2020near,wang2020integrated,pelucchi2022potential}, offering high optical nonlinearity, phase stability, and complementary metal-oxide-semiconductor (CMOS) compatibility. With the advance in this field, integrating a large number of optical elements onto a single silicon chip is feasible, enabling the generation and control of diverse quantum states\cite{wang2016chip,li2017chip,lu2020three,vigliar2021error,appas2021flexible,sharma2022silicon,xia2022experimental,chen2023chip,bao2023very,zheng2023multichip,miloshevsky2024cmos}. These advances support the realization of large-scale quantum information processing and communication, with photonic chip-based entanglement distribution experiments being conducted\cite{appas2021flexible,ren2023photonic,liu2023photonic,jing2024experimental,du2024demonstration,zhao2024long}. 

In this work, we employ a polarization-entangled photon-pair source that generates photon pairs across the entire telecom C-band ($\sim$36 nm), offering arbitrary spectral bandwidth selection. When filtered using dense wavelength-division multiplexing (DWDM) channels with standard 100 GHz bandwidth, the source generates 22 frequency channel pairs for multiplexing schemes\cite{li2017chip,wengerowsky2018entanglement,joshi2018frequency,pseiner2021experimental}. We employ one photon pair after propagating 30 kilometers through single-mode fiber. To ensure accurate measurements of coincidence events, clocks at both sides are synchronized with an accuracy of about 0.322 ns, which is achieved by satellite signals and the intrinsic time correlations of photon pairs in the spontaneous four-wave mixing (SFWM) process\cite{li2005optical,lin2006silicon,sharping2006generation,clemmen2009continuous,chen2021quantum}. Finally, we violate the Clauser-Horne-Shimony-Holt (CHSH) version of Bell’s inequality\cite{clauser1969proposed} by 27.8 standard deviations. Compared to resonant rings\cite{miloshevsky2024cmos}, the nanowires are less susceptible to environmental perturbations, making them robust for practical applications. Our silicon-on-insulator (SOI) platform enables highly integrated, telecom-wavelength compatible polarization-entangled photon-pair sources operating at room temperature. With high generation rates and frequency-multiplexing capacity, this photonic chip offers a scalable solution for distributed quantum computing and fully-connected quantum networks\cite{zheng2023multichip,joshi2020trusted,wen2022realizing,liu202240}.

\section{Experimental Setup}

    \begin{figure*}
      \includegraphics[width=1\linewidth]{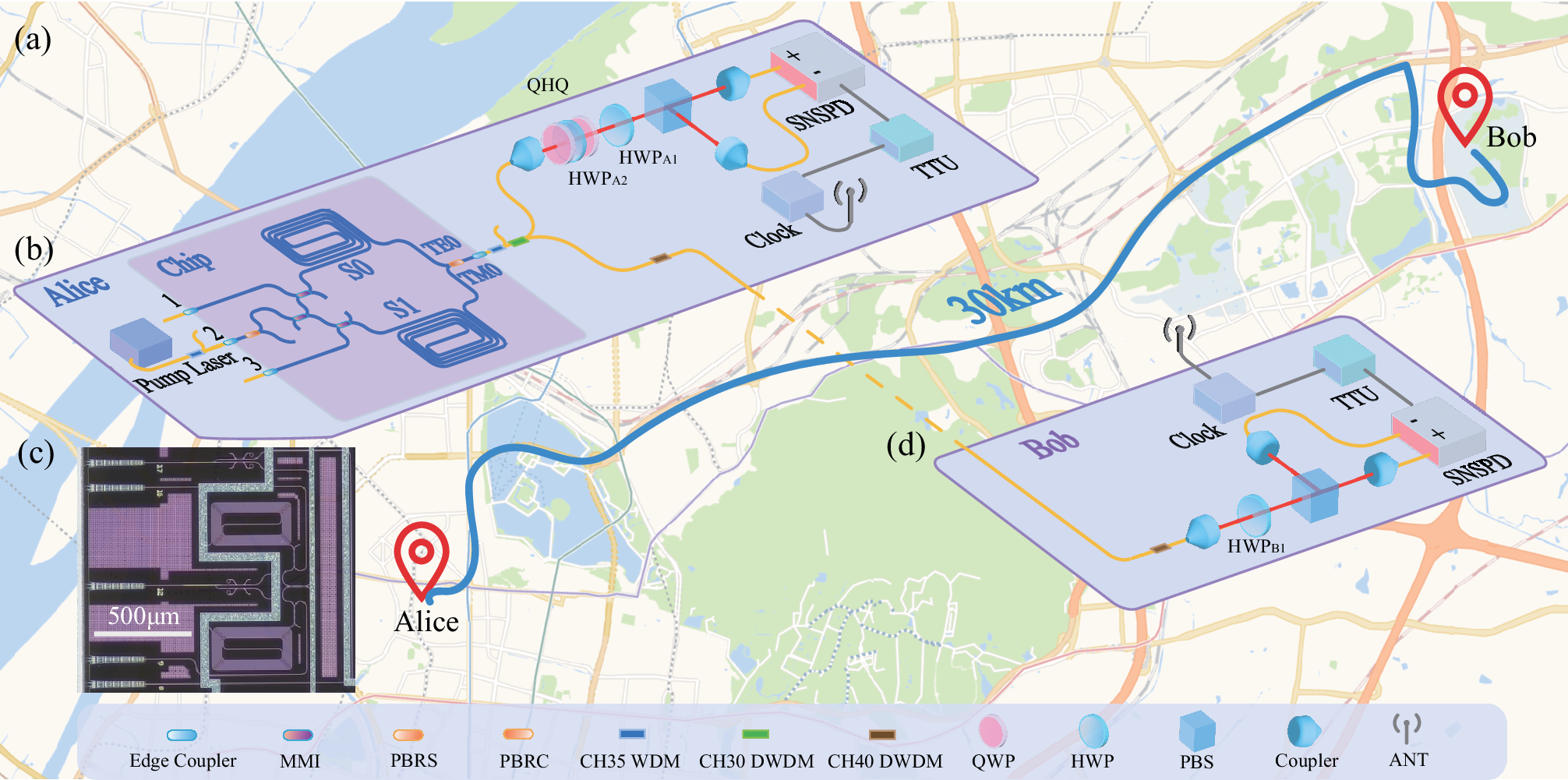}
      
      \caption{\label{Fig1} Network layout and experimental setup. (a) Layout of the two-node quantum network. A 30 km metropolitan optical fiber connects Alice and Bob across the city of Nanjing, China, enabling entanglement distribution. (b) The continuous-wave (CW) pump laser is coupled into the chip via an edge coupler. Laser coupled into edge coupler 2 produces the entangled state, while couplers 1 and 3 enable characterization of individual nanowires. On the chip, the two-photon polarization-entangled state is generated and divided into idler and signal channels using DWDM at channels 30 and 40 of the International Telecommunication Union (ITU) 100 GHz DWDM grid. The idler photon is measured locally by Alice, while the signal photon is transmitted to Bob with the 30 km fiber link. (c) A microscope image of the silicon photonic chip. (d) After 30 km transmission, signal photons are received by Bob, who adjusts the half-wave plate ($HWP_{B1}$) for polarization rotation before performing polarization measurements. Map data from Amap.}
    \end{figure*}

As shown in Fig.~\ref{Fig1}(a), the quantum network consists of two nodes, Alice and Bob, located in the city of Nanjing at the Gulou and Xianlin campuses of Nanjing University, and connected by a 30 km single-mode fiber for entanglement distribution. Fig.~\ref{Fig1}(b) illustrates the generation of polarization-entangled photon pairs on a silicon-on-insulator (SOI) platform, with projective measurements performed on the idler photons. The continuous-wave (CW) pump laser operates at the wavelength of 1549.32 nm, corresponding to channel 35 (CH35) of the International Telecommunication Union’s (ITU) 200 GHz wavelength-division multiplexing (WDM) grid. The pump laser, filtered by WDM at channel 35 (CH35 WDM), is coupled into the SOI chip, which was manufactured at the Advanced Micro Foundry (AMF) with 500 nm×220 nm fully-etched silicon nanowire.

\begin{figure*}
          \includegraphics[width=1\linewidth]{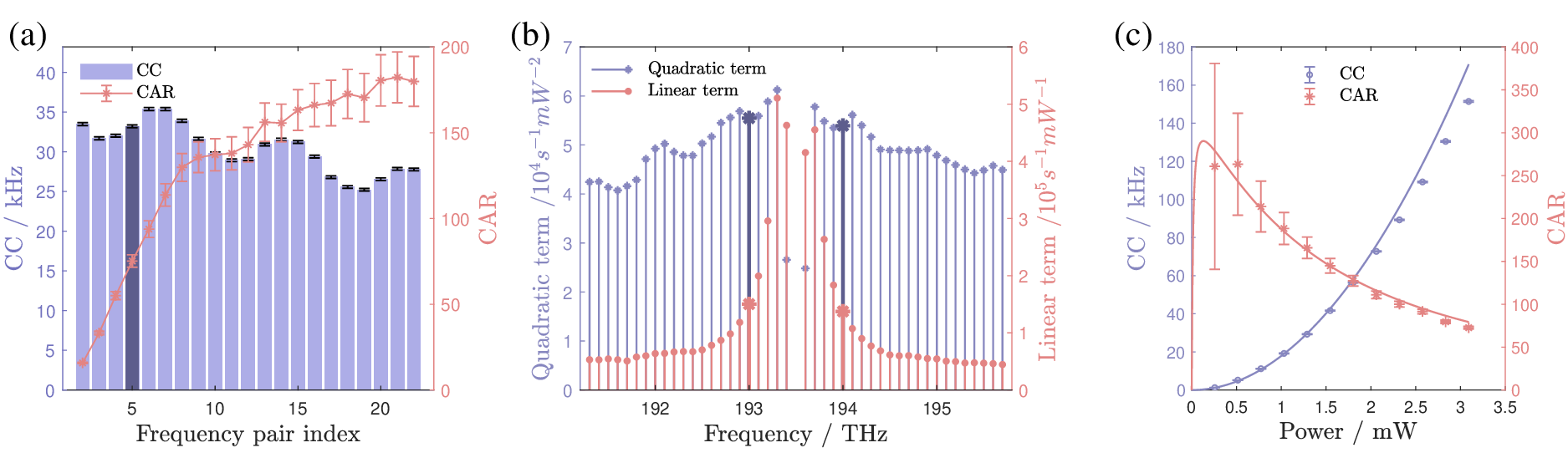}
  	\caption{\label{Fig2} Multiple photon pairs generation from the silicon chip. (a) Coincidence counts (CC) and coincidence-to-accidental ratio (CAR) for all frequency pairs of the two-photon polarization-entangled state. Frequency pair 5 (CH30 and CH40), highlighted in a darker color, is selected for entanglement distribution. A programmable wavelength selective switch (WSS) separates photons within a 100 GHz bandwidth in the C-band. The results presented are obtained after replacing the CH30 and CH40 DWDM modules with the WSS. (b) The quadratic coefficient represents the expected photon production rate via SFWM, while the linear coefficient corresponds to the Raman noise photon production rate. (c) CC and CAR as functions of pump power for frequency pair 5 with 100 GHz DWDMs. Error bars are derived from Poissonian statistics.}
    \end{figure*}   

 Inside the silicon waveguide, the pump propagates in two fundamental modes: the quasi-transverse electric mode (TE0) and the quasi-transverse magnetic mode (TM0). An on-chip polarization beam rotator splitter (PBRS) \cite{sacher2014polarization,chen2016design} separates the TM0 and TE0 modes and converts the TM0 mode into TE0. The TE0-mode pump laser is subsequently divided into two beams by a multimode interferometer (MMI), which then pump entanglement sources based on two spiral nanowires, S0 and S1. The sources S0 and S1 generate path-encoded photon pairs $|00\rangle$ and $|11\rangle$ through SFWM, where energy conservation ensures that the photon frequencies are symmetrically distributed around the pump photons frequency. A polarization beam rotator combiner (PBRC), operating in reverse orientation compared to the PBRS, converts the TE0-mode photons from the S1 nanowire into TM0 mode, and combines them with the TE0-mode photons from the S0 nanowire. This process generates polarization-entangled states entirely on a single chip, indicating the functionality and versatility of integrated quantum photonics. The schematics and layout are respectively shown in Fig.~\ref{Fig1}(b) and \ref{Fig1}(c). The two-photon polarization entangled state output from the chip is:
 \begin{equation}
    \label{eq1}
    \begin{aligned}
        |\Psi\rangle &= \frac{1}{\sqrt{2}} \left( |H\rangle_s |H\rangle_i + e^{i\theta} |V\rangle_s |V\rangle_i \right) \\
        &\otimes \frac{1}{\sqrt{M}} \sum_{m=1}^{M} |\omega\rangle_{s,m} |\omega\rangle_{i,m}.
    \end{aligned}
\end{equation}
 $|H\rangle$ and $|V\rangle$ represent the TE0 and TM0 modes, respectively. $\theta$ represents the relative phase. $|\omega\rangle_{s(i),m}$ represents the quantum state of the signal (idler) photon in frequency channel pair indexed by m (m = 1, 2, \ldots, M).

The CH35 WDM module filters out the pump component from the output light. To isolate idler photons for local measurements, the CH30 DWDM is used, followed by a 'sandwiched' setup (QHQ) consisting of a quarter-wave plate, half-wave plate, and another quarter-wave plate (as shown in Fig.~\ref{Fig1}(b)). This setup compensates for the relative phase $\theta$ between the entangled biphoton states\cite{silverstone2014chip}. $HWP_{A1/B1}$ projects photons onto different polarization measurement bases. Signal photons, filtered by the CH40 DWDM for entanglement distribution, are then sent to the distant Bob for measurement (Fig.~\ref{Fig1}(d)) with about 30 km deployed standard optical fiber, shown in Fig.~1. The relevant detection events on Alice’s and Bob’s sides are recorded with independent time-tagging units each disciplined to either the BeiDou Navigation Satellite System or the Global Positioning System (GPS)\cite{ursin2007entanglement,ma2012quantum,marcikic2006free,scheidl2009feasibility}.

\section{Results}

On Alice’s side, we perform measurements with the integrated polarization-entangled photon pair source, selecting a 100 GHz bandwidth for each channel. Spiral nanowires, 1 cm long with a cross-section of about 220 nm × 500 nm, generate photon pairs with a continuous spectrum. Photon pairs produced by SFWM are extracted by isolating photons symmetrically distributed around the pump laser frequency of 193.5 THz (CH35). A programmable wavelength selective switch (WSS) is used to separate a series of frequency pairs with a 100 GHz bandwidth and to perform fast frequency selection measurements shown in Fig. \ref{Fig2}(a) and (b). 
 
The coincidence counts (CC) shown in Fig. \ref{Fig2}(a) are around 30 kHz for 22 pairs of frequency channels, with an on-chip pump power of 3.09 mW. Significant improvement in CC can be achieved by reducing insertion loss, particularly those from the chip (2.8 dB per facet) and the WSS (4.5 dB). Due to the 200 GHz bandwidth of the CH35 WDM, photons at 193.4 THz and 193.6 THz experience considerable loss, so the first frequency pair is excluded from the analysis. We calculate the Coincidence-to-Accidental Ratio (CAR) using the formula:
\begin{equation}
    \label{eq2}
    CAR = \frac{CC_{true}}{ACC} = \frac{CC - ACC}{ACC},
\end{equation}
where $\mathrm{CC}_\mathrm{true}$ is true coincidence counts, CC is the experimentally measured coincidence counts, ACC is the accidental counts accounting for noise contributions.As the frequency pair moves away from the pump frequency, the coincidence-to-accidental ratio (CAR) increases because the Raman noise from the pump laser affects those pairs near the pump frequency and diminishes with increasing frequency separation. In Fig. \ref{Fig2}(a), the darker data bar represents the CC of frequency pair 5, corresponding to the CH30 and CH40 frequency pair, which is used for entanglement distribution. This pair achieves CC at about 33.2 kHz and CAR at about 75 using the WSS.

We then increase the power and use WSS to scan single counts (SC) from 191.3 THz to 195.7 THz with a step size and bandwidth of 100 GHz. Based on the photon production rate model:
\begin{equation}
    \label{eq3}
    \begin{aligned}
        SC & = \eta \cdot P_{GR} \cdot P^2 + \eta \cdot N_{RS} \cdot P + N_{DC} \\
        & = aP^2 + bP + c,
    \end{aligned}
\end{equation}
we perform fitting to extract the quadratic coefficient $a$ and linear coefficient $b$, where $\eta$ represents the total system efficiency (incorporating coupling, insertion, and detection losses), PGR is the photon-pair generation rate of the SFWM process, P denotes the on-chip power, $\mathrm{N}_\mathrm{RS}$ accounts for Raman scattering noise, and $\mathrm{N}_\mathrm{DC}(c)$ reflects the dark counts\cite{zhang2019generation,jing2024experimental,zhao2024long,wen2022realizing,wen2023polarization,wang2025bright,lin2007photon,reimer2016generation}. The quadratic coefficient, $a$, has values around $5 \times 10^4  \text{s}^{-1}  \mathrm{mW}^{-2}$. The linear coefficient, $b$, is relatively high near 193.5 THz, as shown in Fig. \ref{Fig2}(b), consistent with the lower CAR observed in Fig. \ref{Fig2}(a). According to the quadratic coefficient in Fig. \ref{Fig2}(b) and the actual insertion loss, the average brightness is about $8.7 \times 10^5  \text{s}^{-1}  \text{nm}^{-1}  \mathrm{mW}^{-2}$.

In Fig. \ref{Fig2}(c), after replacing the WSS with the CH30 and CH40 DWDM, we gradually increase the power and measure CC and CAR of frequency pair 5. The quadratic dependence $CC_{true} = \eta_s \eta_i P_{\mathrm{GR}} P^2 = \gamma P^2$ is fitted to extract the coefficient $\gamma$, where $\gamma$ represents the coincidence rate\cite{zhang2019generation}. The fitting yields
$\gamma = 2.23 \times 10^{4}\ \mathrm{s^{-1}\,nm^{-1}\,mW^{-2}}.$
Subsequently, the CAR is characterized using the following expression:
\begin{equation}
    \label{eq4}
    CAR = \frac{CC_{true}}{ACC} = \frac{\gamma P^2}{SC_s SC_i W_t},
\end{equation}
where $\mathrm{SC}_\mathrm{s/i}$ are the single-count rates in the signal/idler channels, and $\mathrm{W}_\mathrm{t}$ is the coincidence detection window for CC\cite{zhang2019generation}.The maximum CC, achieved with an on-chip power of 3.09 mW, reaches about 154 kHz. As the power increases, the CC rises rapidly, but multi-photon events result in a significant increase in accidental coincidence counts, reducing the CAR. Under high-power conditions, multiphoton absorption and free-carrier absorption reduce the actual CC below the fitted values\cite{bristow2007two,schroder1978free}.

\begin{figure}
      \includegraphics[width=1\linewidth]{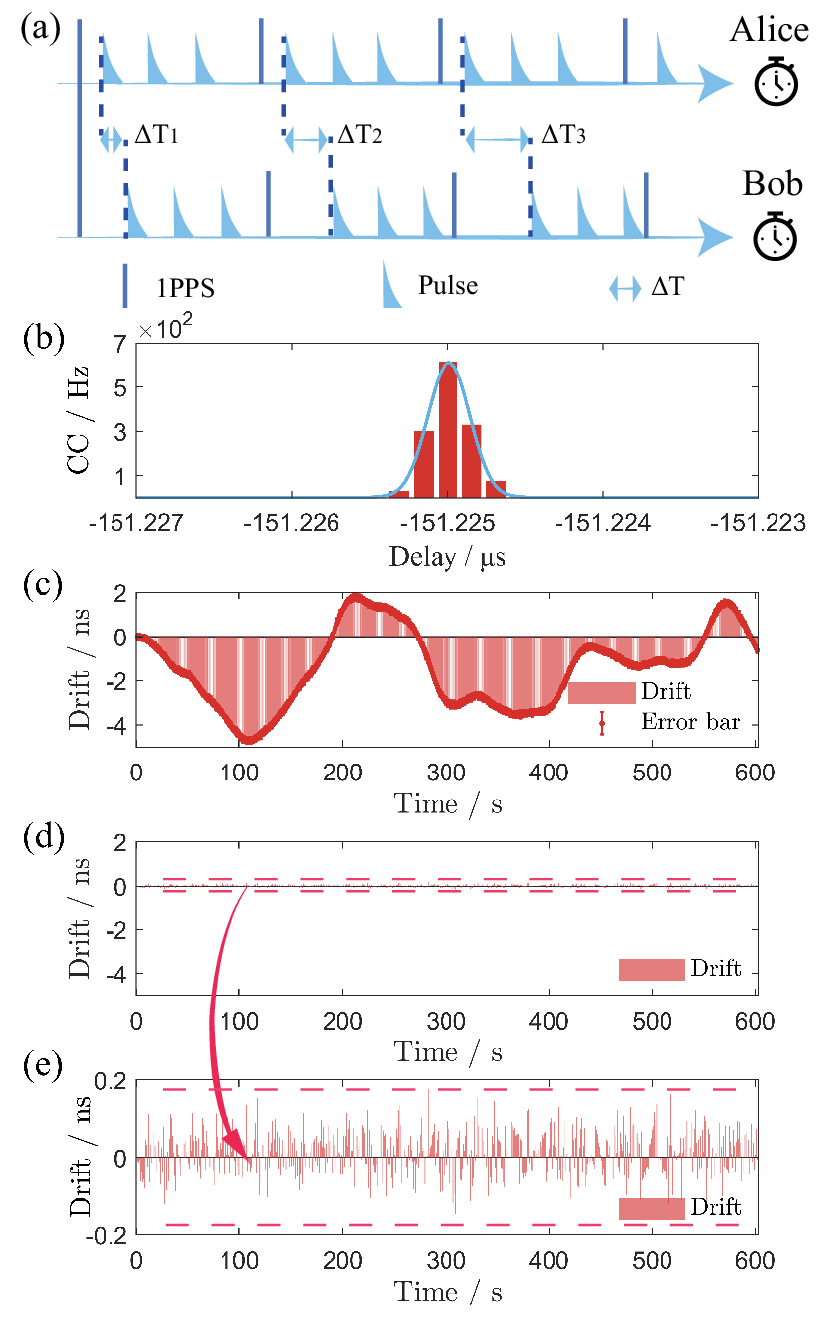}
\caption{\label{Fig3} Clock synchronization in entanglement distribution. (a) Time tags from Alice and Bob are divided into 1-second data blocks based on the 1 pulse per second (1PPS) signal. The time tag alignment is manually set in the first data block. The delay of CC in the i-th block is denoted as $\Delta T_i$. (b)The horizontal axis at the maximum value of CC is the actual delay between Alice and Bob. The results show a time difference of 151.225 $\mu$s, corresponding to the temporal offset introduced by a 30.245 km fiber. The blue line represents the fitted Gaussian function. (c) Red columns represent average delay drifts of CC after aligning the delay of the first second. The time drift is not calibrated by intrinsic time correlations, yielding a standard deviation of 1.7500 ns. (d) Delay results are obtained after calibrating the drift of each second. The standard deviation is suppressed to 0.0568ns. (e) Results after enlarging the vertical axis of Fig. 3(d).}
\end{figure}

After the entanglement distribution with the 30 km deployed fiber connecting Gulou and Xianlin Campuses of Nanjing University, the idler and signal photons are detected by Alice and Bob, respectively. To ensure accurate CC measurements, the local clocks at both sides should be synchronized. By doing so, the drift in time tags between two remote locations can be controlled and minimized. To achieve this, clocks are placed at each location to calibrate the time tagging units (TTU). These clocks are synchronized using satellite signals from either the BeiDou Navigation Satellite System or the GPS, received through individual antennas, and provide 1 Pulse Per Second (1PPS) and 10 MHz signals to the TTU. The 1PPS signal serves as the time reference for Alice and Bob’s time tags, dividing the TTU output into 1 second data blocks. The 10 MHz signal is used to discipline the TTUs’ clocks, ensuring a nominal relative drift of less than $5 \times 10^{-12}  \text{s}^{-1}$.  However, even after this calibration, the drift between two remote clocks can reach up to 100 ns within a few hours. This accuracy is still two orders of magnitude higher than what we need ($\sim0.322$\,ns) . To address this challenge, we further use the intrinsic time correlations in the photon-pair generation process to synchronize the two clocks\cite{ursin2007entanglement,ma2012quantum,marcikic2006free,scheidl2009feasibility}. 

The working principle of our clock synchronization is shown in Fig. \ref{Fig3}(a). We first align the 1PPS signals from Alice and Bob, and $\Delta\mathrm{T}_\mathrm{i}$ represents the actual delay of the time tags in the i-th second after this initial alignment. Scanning the delay in the first second of data (Fig. \ref{Fig3}(b)), the maximum value of CC is found at about 151.225 $\mu$s (=$\Delta\mathrm{T}_\mathrm{1}$), which is the initial offset between Alice and Bob caused by the 30.245 km fiber transmission. The full width at half maximum (FWHM) of the Gaussian function in Fig. \ref{Fig3}(b) is 0.322 ns, which corresponds to the temporal accuracy of our system within the integration time (${\sim}1$\,s). The red columns in Fig. \ref{Fig3}(c) illustrate the overall drift relative to the first second. These drifts, without being calibrated, are equal to the delay minus the initial offset ($\Delta\mathrm{T}_\mathrm{i}-\Delta\mathrm{T}_\mathrm{1}$). The peak-to-peak drift within 600 s is about 6.5 ns. This drift will significantly reduce the CAR from 47.4 to 5.3.

After eliminating the initial offset ($\Delta\mathrm{T}_\mathrm{1}$), we scan the delay in the following second of data to obtain the drift $\Delta\mathrm{t}_\mathrm{2} = \Delta\mathrm{T}_\mathrm{2}-\Delta\mathrm{T}_\mathrm{1}$ relative to the first second, illustrated in Fig. \ref{Fig3}(a). Similarly, after eliminating $\Delta\mathrm{T}_\mathrm{2}$, the delay in the 3rd second of data is scanned to calculate the drift $\Delta\mathrm{t}_\mathrm{3} = \Delta\mathrm{T}_\mathrm{3}-\Delta\mathrm{T}_\mathrm{2}$ relative to the 2nd second, and this process continues iteratively. This iterative method significantly reduces the computational effort. Fig. \ref{Fig3}(d) shows the results after eliminating the overall drift ($ \Delta\mathrm{T}_\mathrm{i}-\Delta\mathrm{T}_\mathrm{1}$), where the residual relative drifts are reduced to below 176 ps in 600 s. This may be limited by the rapid drift of 10 MHz signals and the TTU’s finite temporal resolution of 156 ps. We use standard deviation to quantify the overall drift $ \Delta\mathrm{T}_\mathrm{i}-\Delta\mathrm{T}_\mathrm{1}$ on N blocks of data integrated with 1-second scale.
 
Fig. \ref{Fig3}(e) displays a magnified view of the temporal synchronization results from Fig. \ref{Fig3}(d). The implemented post-processing method reduces the standard deviation from 1.7500 ns to 0.0568 ns, achieving high-quality clock synchronization.

	\begin{figure}
          \includegraphics[width=1\linewidth]{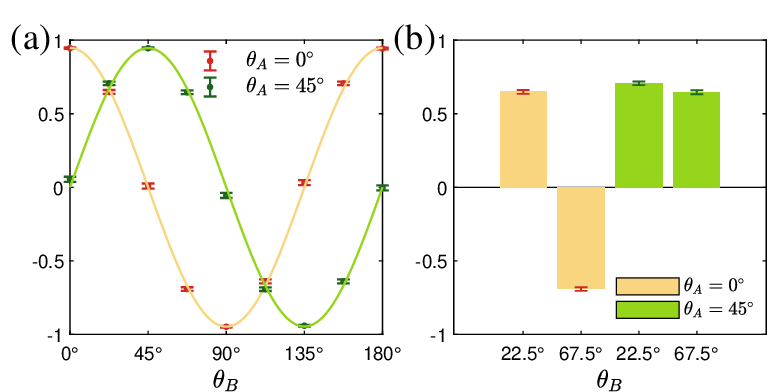}
  	\caption{\label{Fig4} Polarization correlation function of remote entanglement distribution between Alice and Bob. (a) Alice sets $\theta_A$ to 0° (red points, fitted with the yellow curve) and 45° (green points, fitted with the green curve), while Bob sweeps $\theta_B$ from 0° to 180°. (b) Four expectation values, $E(0^\circ, 22.5^\circ)$, $E(0^\circ, 67.5^\circ)$, $E(45^\circ, 22.5^\circ)$, and $E(45^\circ, 67.5^\circ)$, are used to calculate the S parameter of the CHSH inequality. $E_1(0^\circ, 22.5^\circ) = 0.6482 \pm 0.0131$, $E_2(0^\circ, 67.5^\circ) = -0.6917 \pm 0.0123$, $E_3(45^\circ, 22.5^\circ) = 0.7065 \pm 0.0122$, $E_4(45^\circ, 67.5^\circ) = 0.6461 \pm 0.0133$. Error bars are derived from Poissonian statistics and error propagation. }
  	\end{figure}
   
To verify the entanglement of the quantum state after distribution, we demonstrate the violation of the CHSH inequality. After compensating the phase $\theta$ using the QHQ setup, Alice sets $\theta_A$ to 0° and 45°, while Bob sweeps $\theta_B$ from 0° to 180° in steps of 22.5°. We calculate the normalized expectation value $E(\theta_A, \theta_B)$ based on the polarization measurement results obtained by Alice and Bob. The coincidence count rate is approximately 3 kHz for the remote entanglement distribution, and expectation values for each angular combination are calculated using the formula:
    \begin{equation}
		\label{eq5}
  		E(\theta_A, \theta_B) = \frac{CC_{++} + CC_{--} - CC_{+-} - CC_{-+}}{CC_{++} + CC_{--} + CC_{+-} + CC_{-+}},
	\end{equation}
where $ \mathrm{CC}_\mathrm{ij}$ represents the CC of the polarization analyzer’s output, with $\mathrm{i}, \mathrm{j} \in \{+, -\}$ indicating its two output channels, $\theta_A$ and $\theta_B$ are angles of polarization analyzer. The fitted correlation function curves are presented in Fig. \ref{Fig4}(a). Maximum values of $E(\theta_A, \theta_B)$ are 0.9460 ± 0.0055 for $\theta_A = 0^\circ$ and 0.9454 ± 0.0056 for $\theta_A = 45^\circ$. Using four data points at $\theta_A = 0^\circ / 45^\circ$ and $\theta_B = 22.5^\circ / 67.5^\circ$, as shown in Fig. \ref{Fig4}(b), we obtain the S-value of the CHSH inequality:
    \begin{equation}
		\label{eq6}
  		S = \left| E_1 - E_2 + E_3 + E_4 \right| = 2.6926 \pm 0.0249.
	\end{equation}
 
This value exceeds the classical limit of S=2 by 27.8 standard deviations, confirming the successful distribution of entanglement.

\section{Conclusions}
    We have demonstrated an integrated, broadband, polarization-entangled source based on silicon chip, achieving entanglement distribution in a two-node quantum network and successfully violating the CHSH inequality. The on-chip source exhibited high brightness, offering a continuous spectrum spanning 4.5 THz. This broad range allows for flexible channel bandwidth selection and supports multi-user operation, enhancing its versatility. During the entanglement distribution process, we implemented an efficient clock synchronization method based on the intrinsic time correlations in the SFWM process, and violated the CHSH inequality by 27.8 standard deviations. Our work establishes a scalable and integrated platform for the generation of broadband entangled photons, paving the way for the development of advanced multi-user quantum networks and marking a significant step toward practical, scalable quantum communication systems.

\begin{acknowledgments}
This research was supported by the National Key Research and Development Program of China (Grants Nos. 2022YFE0137000), the Natural Science Foundation of Jiangsu Province (Gsrants Nos. BK20240006，BK20233001)), the Innovation Program for Quantum Science and Technology (Grants Nos. 2021ZD0300700 and 2021ZD0301500), and Nanjing University-China Mobile Communications Group Co.,Ltd. Joint Institute.
\end{acknowledgments}

\section*{Data availability}
The data that support the findings of this article are openly available\cite{data_availability}.

\appendix

\section{Characterization of a Single Nanowire }
    
\begin{figure}
          \includegraphics[width=1\linewidth]{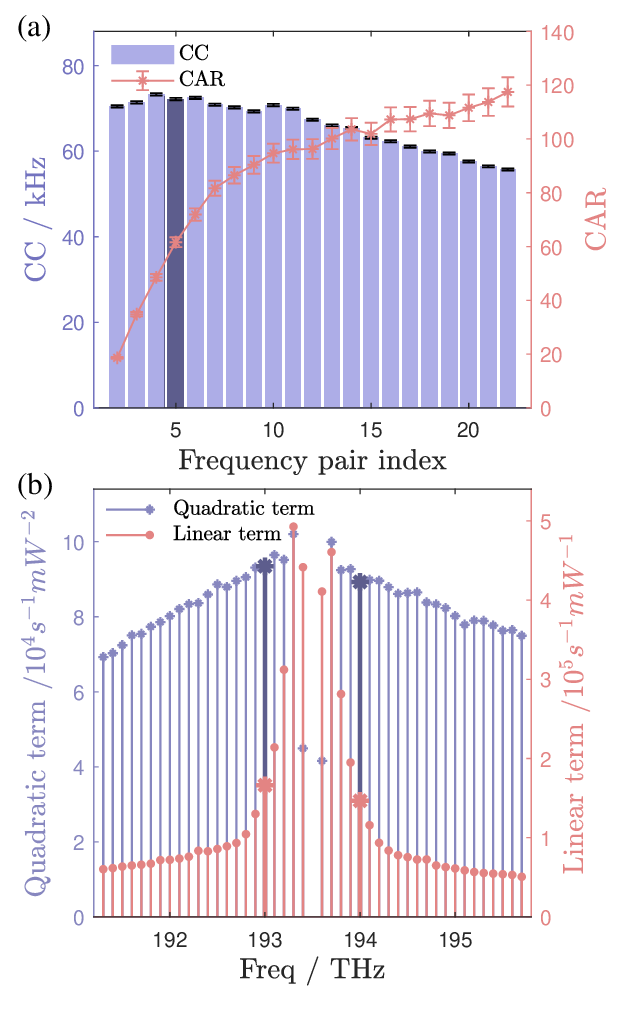}
  	\caption{\label{Fig5} Multiple photon pairs generation from a single silicon nanowire. (a) CC and CAR of the single nanowire for all pairs. (b) The linear and quadratic terms can be separated by fitting each frequency line. The quadratic term coefficient represents the expected photon production rate, while the linear term coefficient represents the Raman noise photon production rate. }
  	\end{figure}

 Fig. \ref{Fig5} characterizes the single nanowire configuration depicted in Fig. \ref{Fig1}(b), where the pump laser is coupled through coupler 1. The photon production rate is proportional to $P^2$, resulting in CC and quadratic term for the single nanowire being twice as high as those of the entangled photon source, while the linear coefficient remains unchanged. These are consistent with the results in Figs. \ref{Fig2} and FIG. \ref{Fig5}.

We also find that the linear term in the low-frequency part is slightly higher than high-frequency part, reflecting the asymmetry of the noise spectrum. This indicates that the Stokes part is larger than the anti-Stokes part\cite{lin2007photon}. As the frequency shifts away from the pump frequency in FIG. \ref{Fig5}(a), the quadratic term decreases gradually due to the influence of anomalous group-delay dispersion\cite{turner2006tailored}. 

\section{On-Chip Device Parameters}

\begin{table}[htbp]
    \centering
    \caption{Loss of each device at 1550nm.}
    \label{tab:loss}
    \begin{tabular*}{\columnwidth}{@{\extracolsep{\fill}}l l c@{}}
    \hline\hline 
    \textbf{Device} & \textbf{Mode} & \textbf{Typical Loss / dB} \\
    \hline
    Waveguide & TE & 2/cm \\
    Edge Coupler & TE & 1.3 \\
    MMI &  & 0.15 \\
    PBRS & TE to TE & 0.1 \\
    PBRS & TM to TE & 0.2 \\
    PBRC & TE to TE & 0.43 \\
    PBRC & TE to TM & 0.25 \\
    \hline\hline 
    \end{tabular*}
\end{table}

\begin{table}[htbp]
    \centering
    \caption{Crosstalk of PBRS at 1550nm.}
    \label{tab:crosstalk}
    \begin{tabular*}{\columnwidth}{@{\extracolsep{\fill}}l l c@{}}
    \hline\hline
    \textbf{Device} & \textbf{Mode} & \textbf{Crosstalk / dB} \\
    \hline
    PBRS & TE to TE & 19.4 \\
    PBRS & TM to TE & 18.5 \\
    \hline\hline
    \end{tabular*}
\end{table}

\section{On-Chip Generation of Polarization-entangled Photon Pairs}

The pump laser coupled into the chip is a coherent superposition of TE0 and TM0 modes of the waveguide. The PBRS routes the TM0-mode to Path B (unused) while directing the TE0 pump laser along Path A. Through adjustment of an off-chip fiber polarization controller, we maximize TE0-mode power in Path A. The PBRS filters out the TM0 mode and optimizes TE0-mode transmission.

A multimode interferometer (MMI) then splits the beam to pump two spiral nanowires, S0 and S1, which generate path-encoded photon pairs $|00\rangle$ and $|11\rangle$, corresponding to $|HH\rangle_0$ and $|HH\rangle_1$, respectively. After passing through a PBRC, $|HH\rangle_0$ remains unchanged, while $|HH\rangle_1$ is converted into $|VV\rangle_1$. The PBRC coherently superposes $|HH\rangle_0$ with $|VV\rangle_1$ along the same path. Finally, a two-photon polarization entangled state is output from the chip: $\frac{1}{\sqrt{2}}(|HH\rangle+|VV\rangle)$.

\begin{figure}
          \includegraphics[width=1\linewidth]{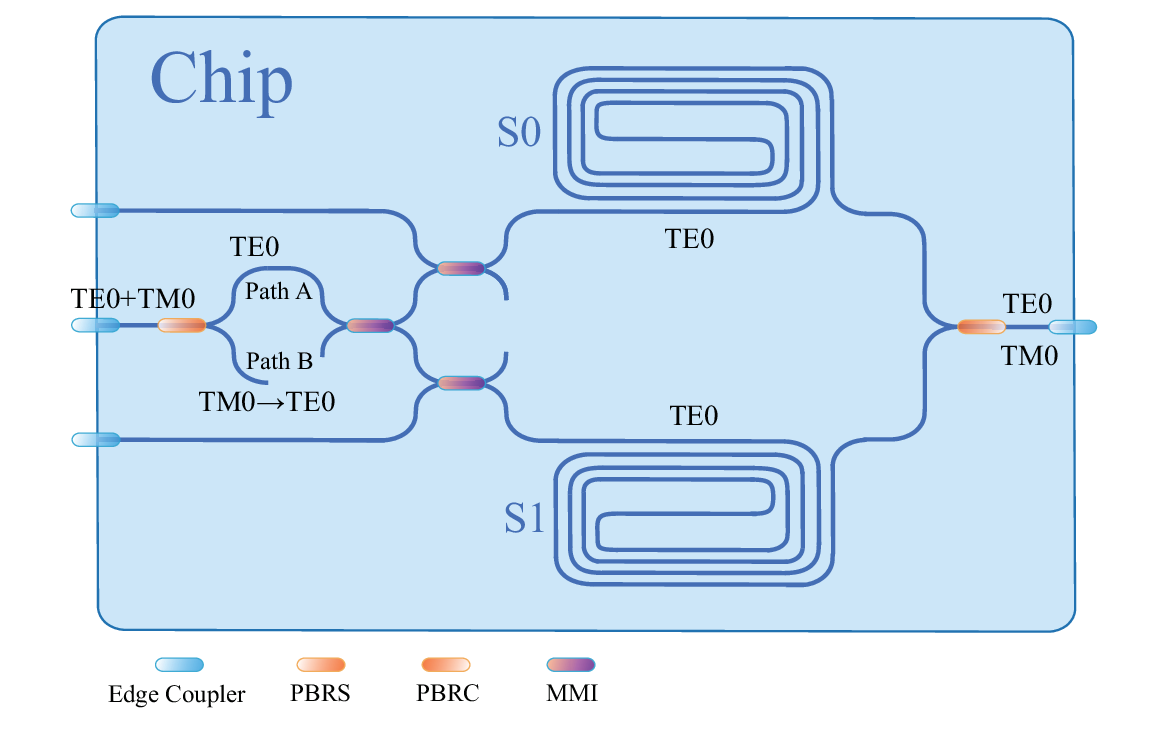}
  	\caption{\label{Fig6} Silicon photonic chip. This chip can generate polarization-entangled photon pairs on-chip, and the image includes various integrated photonic devices.}
  	\end{figure}

 \begin{table*}[htbp]
    \centering
    \caption{Measurement results of CHSH inequality under different conditions.}
    \label{tab:CHSH}
    \begin{tabular*}{\textwidth}{@{\extracolsep{\fill}}l c c c@{}}
    \hline\hline
    {Bandwidth} & {S (CH30\&CH40)} & {S (CH25\&CH45)} & {S (CH20\&CH50)} \\
    \midrule
    100 GHz & $2.5722 \pm 0.0048$ & $2.6909 \pm 0.0045$ & $2.6409 \pm 0.0045$ \\
    200 GHz & $2.5031 \pm 0.0028$ & $2.6311 \pm 0.0033$ & $2.5395 \pm 0.0034$ \\ 
    \hline\hline
    \end{tabular*}
\end{table*}

\begin{table*}[htbp]
    \centering 
    \caption{CC for all channel pairs and bandwidth sets.}
    \label{tab:CC}
    \begin{tabular*}{\textwidth}{@{\extracolsep{\fill}}l c c c@{}}
    \hline\hline
    {Bandwidth} & {CC (CH30\&CH40)} & {CC (CH25\&CH45)} & {CC (CH20\&CH50)} \\
    \midrule
    100 GHz & 102 kHz & 108 kHz & 112 kHz \\
    200 GHz & 271 kHz & 207 kHz & 206 kHz \\ 
    \hline\hline
    \end{tabular*}
\end{table*}

\section{Local CHSH Inequality Measurements}

We perform local CHSH inequality measurements using three different pairs (CH30\&CH40, CH25\&CH45 and CH20\&CH50) of entangled photons with two sets of different bandwidths (100 and 200 GHz). The results in Table \ref{tab:CHSH} all exceed the classical limit of 2. For the 100GHz bandwidth, CC are above 100 kHz, while the 200 GHz bandwidth shows counts above 200 kHz. These results confirm photon-pair entanglement and demonstrate that this broadband chip is suitable for quantum network applications.

The relatively weaker local violations primarily arise from the filtering limitations of the WDMs.

In local CHSH inequality measurements, signal and idler photons generated by the chip are separated using one CH30 WDM and one CH40 WDM, each providing 30 dB noise isolation. Despite this filtering, residual noise photons from other frequency channels persist. Our local measurements yield CHSH values of S=2.5722±0.0048 with one CH40 WDM and S=2.6842±0.0050 with two cascaded CH40 WDMs.

For entanglement distribution, photons transmitted through the 30-km deployed fiber experience significant crosstalk noise from adjacent fibers. This necessitates three cascaded CH40 WDMs for enhanced filtering, suppressing both crosstalk and stray chip-generated photons from other frequency channels. The improved noise suppression achieves stronger CHSH violations in the distributed case (S=2.6926±0.0249), consistent with the local measurements obtained using two WDMs.

\bibliography{Reference.bib}

@article{gisin2007quantum,
  title={Quantum communication},
  author={Gisin, Nicolas and Thew, Rob},
  journal={Nature photonics},
  volume={1},
  number={3},
  pages={165--171},
  year={2007},
  publisher={Nature Publishing Group UK London}
}

@article{ursin2007entanglement,
  title={Entanglement-based quantum communication over 144 km},
  author={Ursin, Rupert and Tiefenbacher, Felix and Schmitt-Manderbach, Tobias and Weier, Henning and Scheidl, Thomas and Lindenthal, Michael and Blauensteiner, Bibiane and Jennewein, Thomas and Perdigues, Josep and Trojek, Pavel and others},
  journal={Nature physics},
  volume={3},
  number={7},
  pages={481--486},
  year={2007},
  publisher={Nature Publishing Group UK London}
}

@article{yin2012quantum,
  title={Quantum teleportation and entanglement distribution over 100-kilometre free-space channels},
  author={Yin, Juan and Ren, Ji-Gang and Lu, He and Cao, Yuan and Yong, Hai-Lin and Wu, Yu-Ping and Liu, Chang and Liao, Sheng-Kai and Zhou, Fei and Jiang, Yan and others},
  journal={Nature},
  volume={488},
  number={7410},
  pages={185--188},
  year={2012},
  publisher={Nature Publishing Group UK London}
}

@article{ma2012quantum,
  title={Quantum teleportation over 143 kilometres using active feed-forward},
  author={Ma, Xiao-Song and Herbst, Thomas and Scheidl, Thomas and Wang, Daqing and Kropatschek, Sebastian and Naylor, William and Wittmann, Bernhard and Mech, Alexandra and Kofler, Johannes and Anisimova, Elena and others},
  journal={Nature},
  volume={489},
  number={7415},
  pages={269--273},
  year={2012},
  publisher={Nature Publishing Group UK London}
}

@article{ren2017ground,
  title={Ground-to-satellite quantum teleportation},
  author={Ren, Ji-Gang and Xu, Ping and Yong, Hai-Lin and Zhang, Liang and Liao, Sheng-Kai and Yin, Juan and Liu, Wei-Yue and Cai, Wen-Qi and Yang, Meng and Li, Li and others},
  journal={Nature},
  volume={549},
  number={7670},
  pages={70--73},
  year={2017},
  publisher={Nature Publishing Group UK London}
}

@article{chen2021integrated,
  title={An integrated space-to-ground quantum communication network over 4,600 kilometres},
  author={Chen, Yu-Ao and Zhang, Qiang and Chen, Teng-Yun and Cai, Wen-Qi and Liao, Sheng-Kai and Zhang, Jun and Chen, Kai and Yin, Juan and Ren, Ji-Gang and Chen, Zhu and others},
  journal={Nature},
  volume={589},
  number={7841},
  pages={214--219},
  year={2021},
  publisher={Nature Publishing Group UK London}
}

@article{lu2022micius,
  title={Micius quantum experiments in space},
  author={Lu, Chao-Yang and Cao, Yuan and Peng, Cheng-Zhi and Pan, Jian-Wei},
  journal={Reviews of Modern Physics},
  volume={94},
  number={3},
  pages={035001},
  year={2022},
  publisher={APS}
}

@book{nielsen2010quantum,
  title={Quantum computation and quantum information},
  author={Nielsen, Michael A and Chuang, Isaac L},
  year={2010},
  publisher={Cambridge university press}
}

@article{ladd2010quantum,
  title={Quantum computers},
  author={Ladd, Thaddeus D and Jelezko, Fedor and Laflamme, Raymond and Nakamura, Yasunobu and Monroe, Christopher and O’Brien, Jeremy Lloyd},
  journal={nature},
  volume={464},
  number={7285},
  pages={45--53},
  year={2010},
  publisher={Nature Publishing Group UK London}
}

@article{lucamarini2018overcoming,
  title={Overcoming the rate--distance limit of quantum key distribution without quantum repeaters},
  author={Lucamarini, Marco and Yuan, Zhiliang L and Dynes, James F and Shields, Andrew J},
  journal={Nature},
  volume={557},
  number={7705},
  pages={400--403},
  year={2018},
  publisher={Nature Publishing Group UK London}
}

@article{wengerowsky2019entanglement,
  title={Entanglement distribution over a 96-km-long submarine optical fiber},
  author={Wengerowsky, S{\"o}ren and Joshi, Siddarth Koduru and Steinlechner, Fabian and Zichi, Julien R and Dobrovolskiy, Sergiy M and Van der Molen, Rene and Los, Johannes WN and Zwiller, Val and Versteegh, Marijn AM and Mura, Alberto and others},
  journal={Proceedings of the National Academy of Sciences},
  volume={116},
  number={14},
  pages={6684--6688},
  year={2019},
  publisher={National Academy of Sciences}
}

@article{dynes2019cambridge,
  title={Cambridge quantum network},
  author={Dynes, JF and Wonfor, Adrian and Tam, WW-S and Sharpe, AW and Takahashi, R and Lucamarini, M and Plews, A and Yuan, ZL and Dixon, AR and Cho, J and others},
  journal={npj Quantum Information},
  volume={5},
  number={1},
  pages={101},
  year={2019},
  publisher={Nature Publishing Group UK London}
}

@article{wengerowsky2020passively,
  title={Passively stable distribution of polarisation entanglement over 192 km of deployed optical fibre},
  author={Wengerowsky, S{\"o}ren and Joshi, Siddarth Koduru and Steinlechner, Fabian and Zichi, Julien R and Liu, Bo and Scheidl, Thomas and Dobrovolskiy, Sergiy M and Molen, Ren{\'e} van der and Los, Johannes WN and Zwiller, Val and others},
  journal={npj Quantum Information},
  volume={6},
  number={1},
  pages={5},
  year={2020},
  publisher={Nature Publishing Group UK London}
}

@article{chen2021twin,
  title={Twin-field quantum key distribution over a 511 km optical fibre linking two distant metropolitan areas},
  author={Chen, Jiu-Peng and Zhang, Chi and Liu, Yang and Jiang, Cong and Zhang, Wei-Jun and Han, Zhi-Yong and Ma, Shi-Zhao and Hu, Xiao-Long and Li, Yu-Huai and Liu, Hui and others},
  journal={Nature Photonics},
  volume={15},
  number={8},
  pages={570--575},
  year={2021},
  publisher={Nature Publishing Group UK London}
}

@article{neumann2022continuous,
  title={Continuous entanglement distribution over a transnational 248 km fiber link},
  author={Neumann, Sebastian Philipp and Buchner, Alexander and Bulla, Lukas and Bohmann, Martin and Ursin, Rupert},
  journal={Nature Communications},
  volume={13},
  number={1},
  pages={6134},
  year={2022},
  publisher={Nature Publishing Group UK London}
}

@article{wang2022twin,
  title={Twin-field quantum key distribution over 830-km fibre},
  author={Wang, Shuang and Yin, Zhen-Qiang and He, De-Yong and Chen, Wei and Wang, Rui-Qiang and Ye, Peng and Zhou, Yao and Fan-Yuan, Guan-Jie and Wang, Fang-Xiang and Chen, Wei and others},
  journal={Nature photonics},
  volume={16},
  number={2},
  pages={154--161},
  year={2022},
  publisher={Nature Publishing Group UK London}
}

@article{liu2023experimental,
  title={Experimental twin-field quantum key distribution over 1000 km fiber distance},
  author={Liu, Yang and Zhang, Wei-Jun and Jiang, Cong and Chen, Jiu-Peng and Zhang, Chi and Pan, Wen-Xin and Ma, Di and Dong, Hao and Xiong, Jia-Min and Zhang, Cheng-Jun and others},
  journal={Physical Review Letters},
  volume={130},
  number={21},
  pages={210801},
  year={2023},
  publisher={APS}
}

@article{takesue2008generation,
  title={Generation of polarization entangled photon pairs using silicon wire waveguide},
  author={Takesue, Hiroki and Fukuda, Hiroshi and Tsuchizawa, Tai and Watanabe, Toshifumi and Yamada, Koji and Tokura, Yasuhiro and Itabashi, Sei-ichi},
  journal={Optics express},
  volume={16},
  number={8},
  pages={5721--5727},
  year={2008},
  publisher={Optical Society of America}
}

@article{leuthold2010nonlinear,
  title={Nonlinear silicon photonics},
  author={Leuthold, Juerg and Koos, Christian and Freude, Wolfgang},
  journal={Nature photonics},
  volume={4},
  number={8},
  pages={535--544},
  year={2010},
  publisher={Nature Publishing Group UK London}
}

@article{silverstone2014chip,
  title={On-chip quantum interference between silicon photon-pair sources},
  author={Silverstone, Joshua W and Bonneau, Damien and Ohira, Kazuya and Suzuki, Nob and Yoshida, Haruhiko and Iizuka, Norio and Ezaki, Mizunori and Natarajan, Chandra M and Tanner, Michael G and Hadfield, Robert H and others},
  journal={Nature Photonics},
  volume={8},
  number={2},
  pages={104--108},
  year={2014},
  publisher={Nature Publishing Group UK London}
}

@article{zhang2019generation,
  title={Generation of multiphoton quantum states on silicon},
  author={Zhang, Ming and Feng, Lan-Tian and Zhou, Zhi-Yuan and Chen, Yang and Wu, Hao and Li, Ming and Gao, Shi-Ming and Guo, Guo-Ping and Guo, Guang-Can and Dai, Dao-Xin and others},
  journal={Light: Science \& Applications},
  volume={8},
  number={1},
  pages={41},
  year={2019},
  publisher={Nature Publishing Group UK London}
}

@article{feng2019generation,
  title={Generation of a frequency-degenerate four-photon entangled state using a silicon nanowire},
  author={Feng, Lan-Tian and Zhang, Ming and Zhou, Zhi-Yuan and Chen, Yang and Li, Ming and Dai, Dao-Xin and Ren, Hong-Liang and Guo, Guo-Ping and Guo, Guang-Can and Tame, Mark and others},
  journal={npj Quantum Information},
  volume={5},
  number={1},
  pages={90},
  year={2019},
  publisher={Nature Publishing Group UK London}
}

@article{paesani2020near,
  title={Near-ideal spontaneous photon sources in silicon quantum photonics},
  author={Paesani, Stefano and Borghi, Massimo and Signorini, Stefano and Ma{\"\i}nos, Alexandre and Pavesi, Lorenzo and Laing, Anthony},
  journal={Nature communications},
  volume={11},
  number={1},
  pages={2505},
  year={2020},
  publisher={Nature Publishing Group UK London}
}

@article{wang2020integrated,
  title={Integrated photonic quantum technologies},
  author={Wang, Jianwei and Sciarrino, Fabio and Laing, Anthony and Thompson, Mark G},
  journal={Nature Photonics},
  volume={14},
  number={5},
  pages={273--284},
  year={2020},
  publisher={Nature Publishing Group UK London}
}

@article{pelucchi2022potential,
  title={The potential and global outlook of integrated photonics for quantum technologies},
  author={Pelucchi, Emanuele and Fagas, Giorgos and Aharonovich, Igor and Englund, Dirk and Figueroa, Eden and Gong, Qihuang and Hannes, H{\"u}bel and Liu, Jin and Lu, Chao-Yang and Matsuda, Nobuyuki and others},
  journal={Nature Reviews Physics},
  volume={4},
  number={3},
  pages={194--208},
  year={2022},
  publisher={Nature Publishing Group UK London}
}

@article{wang2016chip,
  title={Chip-to-chip quantum photonic interconnect by path-polarization interconversion},
  author={Wang, Jianwei and Bonneau, Damien and Villa, Matteo and Silverstone, Joshua W and Santagati, Raffaele and Miki, Shigehito and Yamashita, Taro and Fujiwara, Mikio and Sasaki, Masahide and Terai, Hirotaka and others},
  journal={Optica},
  volume={3},
  number={4},
  pages={407--413},
  year={2016},
  publisher={Optical Society of America}
}

@article{li2017chip,
  title={On-chip multiplexed multiple entanglement sources in a single silicon nanowire},
  author={Li, Yin-Hai and Zhou, Zhi-Yuan and Feng, Lan-Tian and Fang, Wen-Tan and Liu, Shi-long and Liu, Shi-Kai and Wang, Kai and Ren, Xi-Feng and Ding, Dong-Sheng and Xu, Li-Xin and others},
  journal={Physical Review Applied},
  volume={7},
  number={6},
  pages={064005},
  year={2017},
  publisher={APS}
}

@article{lu2020three,
  title={Three-dimensional entanglement on a silicon chip},
  author={Lu, Liangliang and Xia, Lijun and Chen, Zhiyu and Chen, Leizhen and Yu, Tonghua and Tao, Tao and Ma, Wenchao and Pan, Ying and Cai, Xinlun and Lu, Yanqing and others},
  journal={npj Quantum Information},
  volume={6},
  number={1},
  pages={30},
  year={2020},
  publisher={Nature Publishing Group UK London}
}

@article{vigliar2021error,
  title={Error-protected qubits in a silicon photonic chip},
  author={Vigliar, Caterina and Paesani, Stefano and Ding, Yunhong and Adcock, Jeremy C and Wang, Jianwei and Morley-Short, Sam and Bacco, Davide and Oxenl{\o}we, Leif K and Thompson, Mark G and Rarity, John G and others},
  journal={Nature Physics},
  volume={17},
  number={10},
  pages={1137--1143},
  year={2021},
  publisher={Nature Publishing Group UK London}
}

@article{appas2021flexible,
  title={Flexible entanglement-distribution network with an AlGaAs chip for secure communications},
  author={Appas, F{\'e}licien and Baboux, Florent and Amanti, Maria I and Lema{\'\i}tre, Aristide and Boitier, Fabien and Diamanti, Eleni and Ducci, Sara},
  journal={npj Quantum Information},
  volume={7},
  number={1},
  pages={118},
  year={2021},
  publisher={Nature Publishing Group UK London}
}

@article{sharma2022silicon,
  title={Silicon photonic wires for broadband polarization entanglement at telecommunication wavelengths},
  author={Sharma, Shivani and Venkataraman, Vivek and Ghosh, Joyee},
  journal={Physical Review Applied},
  volume={18},
  number={4},
  pages={044043},
  year={2022},
  publisher={APS}
}

@article{xia2022experimental,
  title={Experimental optimal verification of three-dimensional entanglement on a silicon chip},
  author={Xia, Lijun and Lu, Liangliang and Wang, Kun and Jiang, Xinhe and Zhu, Shining and Ma, Xiaosong},
  journal={New Journal of Physics},
  volume={24},
  number={9},
  pages={095002},
  year={2022},
  publisher={IOP Publishing}
}

@article{chen2023chip,
  title={On-chip generation and collectively coherent control of the superposition of the whole family of Dicke states},
  author={Chen, Leizhen and Lu, Liangliang and Xia, Lijun and Lu, Yanqing and Zhu, Shining and Ma, Xiao-song},
  journal={Physical Review Letters},
  volume={130},
  number={22},
  pages={223601},
  year={2023},
  publisher={APS}
}

@article{bao2023very,
  title={Very-large-scale integrated quantum graph photonics},
  author={Bao, Jueming and Fu, Zhaorong and Pramanik, Tanumoy and Mao, Jun and Chi, Yulin and Cao, Yingkang and Zhai, Chonghao and Mao, Yifei and Dai, Tianxiang and Chen, Xiaojiong and others},
  journal={Nature Photonics},
  volume={17},
  number={7},
  pages={573--581},
  year={2023},
  publisher={Nature Publishing Group UK London}
}

@article{zheng2023multichip,
  title={Multichip multidimensional quantum networks with entanglement retrievability},
  author={Zheng, Yun and Zhai, Chonghao and Liu, Dajian and Mao, Jun and Chen, Xiaojiong and Dai, Tianxiang and Huang, Jieshan and Bao, Jueming and Fu, Zhaorong and Tong, Yeyu and others},
  journal={Science},
  volume={381},
  number={6654},
  pages={221--226},
  year={2023},
  publisher={American Association for the Advancement of Science}
}

@article{miloshevsky2024cmos,
  title={CMOS photonic integrated source of broadband polarization-entangled photons},
  author={Miloshevsky, Alexander and Cohen, Lucas M and Myilswamy, Karthik V and Alshowkan, Muneer and Fatema, Saleha and Lu, Hsuan-Hao and Weiner, Andrew M and Lukens, Joseph M},
  journal={Optica Quantum},
  volume={2},
  number={4},
  pages={254--259},
  year={2024},
  publisher={Optica Publishing Group}
}

@article{ren2023photonic,
  title={Photonic-chip-based dense entanglement distribution},
  author={Ren, Shang-Yu and Wang, Wei-Qiang and Cheng, Yu-Jie and Huang, Long and Du, Bing-Zheng and Zhao, Wei and Guo, Guang-Can and Feng, Lan-Tian and Zhang, Wen-Fu and Ren, Xi-Feng},
  journal={PhotoniX},
  volume={4},
  number={1},
  pages={12},
  year={2023},
  publisher={Springer}
}

@article{liu2023photonic,
  title={Photonic-reconfigurable entanglement distribution network based on silicon quantum photonics},
  author={Liu, Dongning and Liu, Jingyuan and Ren, Xiaosong and Feng, Xue and Liu, Fang and Cui, Kaiyu and Huang, Yidong and Zhang, Wei},
  journal={Photonics Research},
  volume={11},
  number={7},
  pages={1314--1325},
  year={2023},
  publisher={Chinese Laser Press and Optica Publishing Group}
}

@article{jing2024experimental,
  title={Experimental quantum Byzantine agreement on a three-user quantum network with integrated photonics},
  author={Jing, Xu and Qian, Cheng and Weng, Chen-Xun and Li, Bing-Hong and Chen, Zhe and Wang, Chen-Quan and Tang, Jie and Gu, Xiao-Wen and Kong, Yue-Chan and Chen, Tang-Sheng and others},
  journal={Science Advances},
  volume={10},
  number={34},
  pages={eadp2877},
  year={2024},
  publisher={American Association for the Advancement of Science}
}

@article{du2024demonstration,
  title={Demonstration of entanglement distribution over 155 km metropolitan fiber using a silicon nanophotonic chip},
  author={Du, Jinyi and Zhang, Xingjian and Chen, George FR and Gao, Hongwei and Tan, Dawn TH and Ling, Alexander},
  journal={arXiv preprint arXiv:2409.17558},
  year={2024}
}

@article{zhao2024long,
  title={Long-distance distribution of telecom time-energy entanglement generated on a silicon chip},
  author={Zhao, Yuan-yuan and Yue, Fuyong and Gao, Feng and Wang, Qibing and Li, Chao and Liu, Zichen and Wang, Lei and He, Zhixue},
  journal={arXiv preprint arXiv:2407.21305},
  year={2024}
}

@article{wengerowsky2018entanglement,
  title={An entanglement-based wavelength-multiplexed quantum communication network},
  author={Wengerowsky, S{\"o}ren and Joshi, Siddarth Koduru and Steinlechner, Fabian and H{\"u}bel, Hannes and Ursin, Rupert},
  journal={Nature},
  volume={564},
  number={7735},
  pages={225--228},
  year={2018},
  publisher={Nature Publishing Group UK London}
}

@article{joshi2018frequency,
  title={Frequency multiplexing for quasi-deterministic heralded single-photon sources},
  author={Joshi, Chaitali and Farsi, Alessandro and Clemmen, St{\'e}phane and Ramelow, Sven and Gaeta, Alexander L},
  journal={Nature communications},
  volume={9},
  number={1},
  pages={847},
  year={2018},
  publisher={Nature Publishing Group UK London}
}

@article{pseiner2021experimental,
  title={Experimental wavelength-multiplexed entanglement-based quantum cryptography},
  author={Pseiner, Johannes and Achatz, Lukas and Bulla, Lukas and Bohmann, Martin and Ursin, Rupert},
  journal={Quantum Science and Technology},
  volume={6},
  number={3},
  pages={035013},
  year={2021},
  publisher={IOP Publishing}
}

@article{li2005optical,
  title={Optical-fiber source of polarization-entangled photons in the 1550 nm telecom band},
  author={Li, Xiaoying and Voss, Paul L and Sharping, Jay E and Kumar, Prem},
  journal={Physical review letters},
  volume={94},
  number={5},
  pages={053601},
  year={2005},
  publisher={APS}
}

@article{lin2006silicon,
  title={Silicon waveguides for creating quantum-correlated photon pairs},
  author={Lin, Q and Agrawal, Govind P},
  journal={Optics letters},
  volume={31},
  number={21},
  pages={3140--3142},
  year={2006},
  publisher={Optical Society of America}
}

@article{sharping2006generation,
  title={Generation of correlated photons in nanoscale silicon waveguides},
  author={Sharping, Jay E and Lee, Kim Fook and Foster, Mark A and Turner, Amy C and Schmidt, Bradley S and Lipson, Michal and Gaeta, Alexander L and Kumar, Prem},
  journal={Optics express},
  volume={14},
  number={25},
  pages={12388--12393},
  year={2006},
  publisher={Optical Society of America}
}

@article{clemmen2009continuous,
  title={Continuous wave photon pair generation in silicon-on-insulator waveguides and ring resonators},
  author={Clemmen, St{\'e}phane and Huy, K Phan and Bogaerts, Wim and Baets, Roel G and Emplit, Ph and Massar, Serge},
  journal={Optics express},
  volume={17},
  number={19},
  pages={16558--16570},
  year={2009},
  publisher={Optical Society of America}
}

@article{chen2021quantum,
  title={Quantum entanglement on photonic chips: a review},
  author={Chen, Xiaojiong and Fu, Zhaorong and Gong, Qihuang and Wang, Jianwei},
  journal={Advanced Photonics},
  volume={3},
  number={6},
  pages={064002--064002},
  year={2021},
  publisher={Society of Photo-Optical Instrumentation Engineers}
}

@article{clauser1969proposed,
  title={Proposed experiment to test local hidden-variable theories},
  author={Clauser, John F and Horne, Michael A and Shimony, Abner and Holt, Richard A},
  journal={Physical review letters},
  volume={23},
  number={15},
  pages={880},
  year={1969},
  publisher={APS}
}

@article{joshi2020trusted,
  title={A trusted node--free eight-user metropolitan quantum communication network},
  author={Joshi, Siddarth Koduru and Aktas, Djeylan and Wengerowsky, S{\"o}ren and Lon{\v{c}}ari{\'c}, Martin and Neumann, Sebastian Philipp and Liu, Bo and Scheidl, Thomas and Lorenzo, Guillermo Curr{\'a}s and Samec, {\v{Z}}eljko and Kling, Laurent and others},
  journal={Science advances},
  volume={6},
  number={36},
  pages={eaba0959},
  year={2020},
  publisher={American Association for the Advancement of Science}
}

@article{wen2022realizing,
  title={Realizing an entanglement-based multiuser quantum network with integrated photonics},
  author={Wen, Wenjun and Chen, Zhiyu and Lu, Liangliang and Yan, Wenhan and Xue, Wenyi and Zhang, Peiyu and Lu, Yanqing and Zhu, Shining and Ma, Xiao-song},
  journal={Physical Review Applied},
  volume={18},
  number={2},
  pages={024059},
  year={2022},
  publisher={APS}
}

@article{liu202240,
  title={40-user fully connected entanglement-based quantum key distribution network without trusted node},
  author={Liu, Xu and Liu, Jingyuan and Xue, Rong and Wang, Heqing and Li, Hao and Feng, Xue and Liu, Fang and Cui, Kaiyu and Wang, Zhen and You, Lixing and others},
  journal={PhotoniX},
  volume={3},
  number={1},
  pages={2},
  year={2022},
  publisher={Springer}
}

@article{sacher2014polarization,
  title={Polarization rotator-splitters in standard active silicon photonics platforms},
  author={Sacher, Wesley D and Barwicz, Tymon and Taylor, Benjamin JF and Poon, Joyce KS},
  journal={Optics express},
  volume={22},
  number={4},
  pages={3777--3786},
  year={2014},
  publisher={Optical Society of America}
}

@article{chen2016design,
  title={Design of an ultra-broadband and fabrication-tolerant silicon polarization rotator splitter with SiO\_2 top cladding},
  author={Chen, Xin and Qiu, Chao and Sheng, Zhen and Wu, Aimin and Huang, Haiyang and Zhao, Yingxuan and Li, Wei and Wang, Xi and Zou, Shichang and Gan, Fuwan},
  journal={Chinese Optics Letters},
  volume={14},
  number={8},
  pages={081301},
  year={2016},
  publisher={OSA}
}

@article{marcikic2006free,
  title={Free-space quantum key distribution with entangled photons},
  author={Marcikic, Ivan and Lamas-Linares, Antia and Kurtsiefer, Christian},
  journal={Applied Physics Letters},
  volume={89},
  number={10},
  year={2006},
  publisher={AIP Publishing}
}

@article{scheidl2009feasibility,
  title={Feasibility of 300 km quantum key distribution with entangled states},
  author={Scheidl, Thomas and Ursin, Rupert and Fedrizzi, Alessandro and Ramelow, Sven and Ma, Xiao-Song and Herbst, Thomas and Prevedel, Robert and Ratschbacher, Lothar and Kofler, Johannes and Jennewein, Thomas and others},
  journal={New Journal of Physics},
  volume={11},
  number={8},
  pages={085002},
  year={2009},
  publisher={IOP Publishing}
}

@article{wen2023polarization,
  title={Polarization-entangled quantum frequency comb from a silicon nitride microring resonator},
  author={Wen, Wenjun and Yan, Wenhan and Lu, Chi and Lu, Liangliang and Wu, Xiaoyu and Lu, Yanqing and Zhu, Shining and Ma, Xiao-Song},
  journal={Physical Review Applied},
  volume={20},
  number={6},
  pages={064032},
  year={2023},
  publisher={APS}
}

@article{wang2025bright,
  title={Bright Heralded Single-Photon Source Saturating Theoretical Single-photon Purity},
  author={Wang, Haoyang and Yuan, Huihong and Zeng, Qiang and Zhou, Lai and Ma, Haiqiang and Yuan, Zhiliang},
  journal={Laser \& Photonics Reviews},
  volume={19},
  number={9},
  pages={2401420},
  year={2025},
  publisher={Wiley Online Library}
}

@article{lin2007photon,
  title={Photon-pair generation in optical fibers through four-wave mixing: Role of Raman scattering and pump polarization},
  author={Lin, Q and Yaman, F and Agrawal, Govind P},
  journal={Physical Review A—Atomic, Molecular, and Optical Physics},
  volume={75},
  number={2},
  pages={023803},
  year={2007},
  publisher={APS}
}

@article{reimer2016generation,
  title={Generation of multiphoton entangled quantum states by means of integrated frequency combs},
  author={Reimer, Christian and Kues, Michael and Roztocki, Piotr and Wetzel, Benjamin and Grazioso, Fabio and Little, Brent E and Chu, Sai T and Johnston, Tudor and Bromberg, Yaron and Caspani, Lucia and others},
  journal={Science},
  volume={351},
  number={6278},
  pages={1176--1180},
  year={2016},
  publisher={American Association for the Advancement of Science}
}

@article{bristow2007two,
  title={Two-photon absorption and Kerr coefficients of silicon for 850--2200nm},
  author={Bristow, Alan D and Rotenberg, Nir and Van Driel, Henry M},
  journal={Applied physics letters},
  volume={90},
  number={19},
  year={2007},
  publisher={AIP Publishing}
}

@article{schroder1978free,
  title={Free carrier absorption in silicon},
  author={Schroder, Dieter K and Thomas, R Noel and Swartz, John C},
  journal={IEEE Journal of solid-state circuits},
  volume={13},
  number={1},
  pages={180--187},
  year={1978},
  publisher={IEEE}
}

@misc{data_availability,
  title={https://github.com/NJU-Malab/SOI-Entanglement-Distribution},
  author={Jiang, Zi-Heng and Yan, Wenhan and Lu, Chi and Chen, Yikai and Wen, Wenjun and An, Yu-Yang and Chen, Leizhen and Liu, Yuchen and Liu, Huaying and Xie, Zhenda and Lu, Yanqing and Zhu, Shining and Ma, Xiao-Song},
  publisher = {GitHub},
  url = {https://github.com/NJU-Malab/SOI-Entanglement-Distribution}, 
  urldate = {2025-08-15},
  year={2025}
}

@article{turner2006tailored,
  title={Tailored anomalous group-velocity dispersion in silicon channel waveguides},
  author={Turner, Amy C and Manolatou, Christina and Schmidt, Bradley S and Lipson, Michal and Foster, Mark A and Sharping, Jay E and Gaeta, Alexander L},
  journal={Optics express},
  volume={14},
  number={10},
  pages={4357--4362},
  year={2006},
  publisher={Optical Society of America}
}

\end{document}